\documentclass[journal,twoside,web]{article}
\usepackage{cite}
\usepackage{amsmath,amssymb,amsfonts}
\usepackage{algorithmic}
\usepackage{graphicx}
\usepackage{algorithm,algorithmic}
\usepackage{hyperref}
\hypersetup{hidelinks=true}
\usepackage{textcomp}
\usepackage{threeparttable}
\usepackage{multirow}
\usepackage{booktabs}

\begin{document}
\title{Cascaded Self-supervised Learning for Subject-independent EEG-based Emotion Recognition}

\author{
Hanqi Wang, Tao Chen, and Liang Song
\thanks{This research is partly funded by the China Mobile Research Fund of Chinese Ministry of Education (Grant No. KEH2310029). This work is also supported by the Shanghai Key Research Lab of NSAI, and Joint Lab on Networked AI Edge Computing Fudan University-Changan. (Corresponding Author: Liang Song, E-mail: songl@fudan.edu.cn)}
\thanks{Hanqi Wang, and Liang Song are with the Academy for Engineering and Technology, Fudan University, Shanghai 200433, China, E-mail:wanghq21@m.fudan.edu.cn.}
\thanks{Tao Chen is with School of Information Science and Technology, Fudan University, Shanghai 200433, China.}
}

\maketitle

\begin{abstract}
EEG-based Emotion recognition holds significant promise for applications in human-computer interaction, medicine, and neuroscience. While deep learning has shown potential in this field, current approaches usually rely on large-scale high-quality labeled datasets, limiting the performance of deep learning. Self-supervised learning offers a solution by automatically generating labels, but its inter-subject generalizability remains under-explored. For this reason, our interest lies in offering a self-supervised learning paradigm with better inter-subject generalizability. Inspired by recent efforts in combining low-level and high-level tasks in deep learning, we propose a cascaded self-supervised architecture for EEG emotion recognition. Then, we introduce a low-level task, time-to-frequency reconstruction (TFR). This task leverages the inherent time-frequency relationship in EEG signals. Our architecture integrates it with the high-level contrastive learning modules, performing self-supervised learning for EEG-based emotion recognition. Experiment on DEAP and DREAMER datasets demonstrates superior performance of our method over similar works. The outcome results also highlight the indispensability of the TFR task and the robustness of our method to label scarcity, validating the effectiveness of the proposed method. 
\end{abstract}


\section{Introduction}
Emotion recognition plays a pivotal role in human-computer interaction and finds widespread applications in fields such as medicine and neuroscience, rendering research in this domain of significant importance. Compared with other techniques, electroencephalography (EEG) has garnered attention as an emotion measurement in this area due to its objectivity and high temporal resolution. Recently, deep learning has been introduced to recognize emotion from EEG signals, achieving notable success~\cite{r4,r5,r6}. However, prevailing methods in deep learning-based EEG emotion recognition are based on supervised learning that needs the label information to guide the training of model~\cite{r0}. Thus, the large-scale labeled training data is generally indispensable for the performance of the supervised model. Meanwhile, the manual annotation of EEG signal labels is time-consuming and laborious, presenting a challenge in collecting large-scale labeled emotional EEG samples~\cite{r0,r2,r3}. In addition, considering that the labeling quality usually relies on expertise and self-report, the collected labels are susceptible to noise and subjective bias that mislead the learned representation of supervised model~\cite{r0,r2,r3}. These facts constrain the performance of the deep learning model in the EEG-based emotion recognition area.

To alleviate the problem, researchers introduce self-supervised learning to extract representation from EEG signals~\cite{r2,r3,r7,r8}. In general, self-supervised learning can automatically generate labels for unlabeled data by constructing a pretext task. This training paradigm mitigates the dependence of the deep learning model on large-scale labeled training data~\cite{r0}. And, the learned comprehensive representation of the underlying structure of the data can serve for following downstream tasks~\cite{r0}. Some recent works have shown the effectiveness of self-supervised learning for EEG emotion recognition~\cite{r2,r3,r7}. However, these advances remain insufficient to cope with the challenges in the field of EEG emotion recognition. The current EEG-based self-supervised learning research pays little attention to improving the generalization of the model on unseen subjects, limiting its practicability in real-life applications. Due to the high inter-subject variability, the EEG signal presents a challenge for deep learning models to generalize across subjects. There is a large discrepancy in the data distribution of EEG signals collected from different subjects. Thus, the subject-dependent model trained on some subjects usually performs inadequately on other unseen subjects, leading to limited performance on subject-independent emotion recognition task. Although some researchers propose to adopt contrastive learning to improve the generalizability of self-supervised learning model~\cite{r7}, a single task learning is insufficient to capture comprehensive subject-invariant information for emotion recognition. Currently, this challenge for self-supervised learning remains under-explore in the field of EEG emotion recognition. Motivated by this, our interest lies in exploring how to improve the generalization ability of the self-supervised EEG emotion recognition model. 

\begin{figure}[ht]
  \centering
  \includegraphics[width=\linewidth]{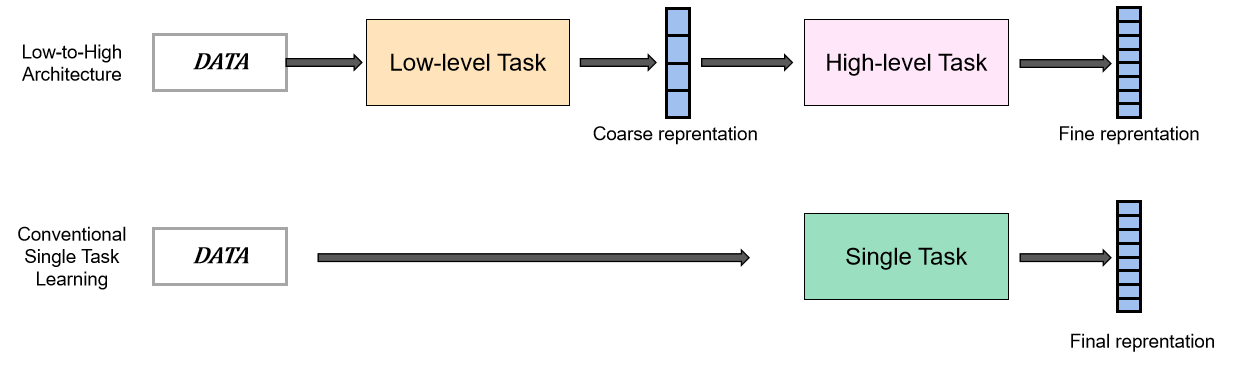}
  \caption{The illustration of cascaded low-to-high architecture and single task learning. Instead of a single task being responsible for the entire learning process, low-to-high architecture evolves from low-level to high-level tasks, pruning the learned representations.}
  \label{fig00}
\end{figure}

Recently, there has been an effort in the deep learning field to combine a low-level task with a high-level task. Usually, the low-level task refers to a task that focuses on the coarse pattern in raw data, while the high-level task involves complex semantic information understanding and reasoning. In some works~\cite{r1,r1-2,r1-3,r1-4,r1-5}, researchers establish a low-to-high cascaded pipeline that jointly optimizes multiple tasks at different levels. In such a pipeline, the cascaded architecture can enable the representation learning to perform coarse-to-fine pruning at different stages, evolving from low-level to high-level as shown in Fig~\ref{fig00}. Empirical findings from these works have demonstrated that connecting a low-level task to a high-level task can improve the generalization ability of trained model~\cite{r1,r1-2,r1-3}. This observation suggests a possible approach to develop a more generalizable method. We hope the low-level task can help capture the subject-invariant simple pattern in the raw EEG data, aligning the distribution of coarse representation of various subjects as shown in Fig~\ref{fig01}. Based on that, the following high-level task continues to refine the extracted coarse representation for the final representation with complex semantic information, improving the generalized capacity. However, the identification of a suitable low-level task remains unexplored in this area.

\begin{figure}[ht]
  \centering
  \includegraphics[width=\linewidth]{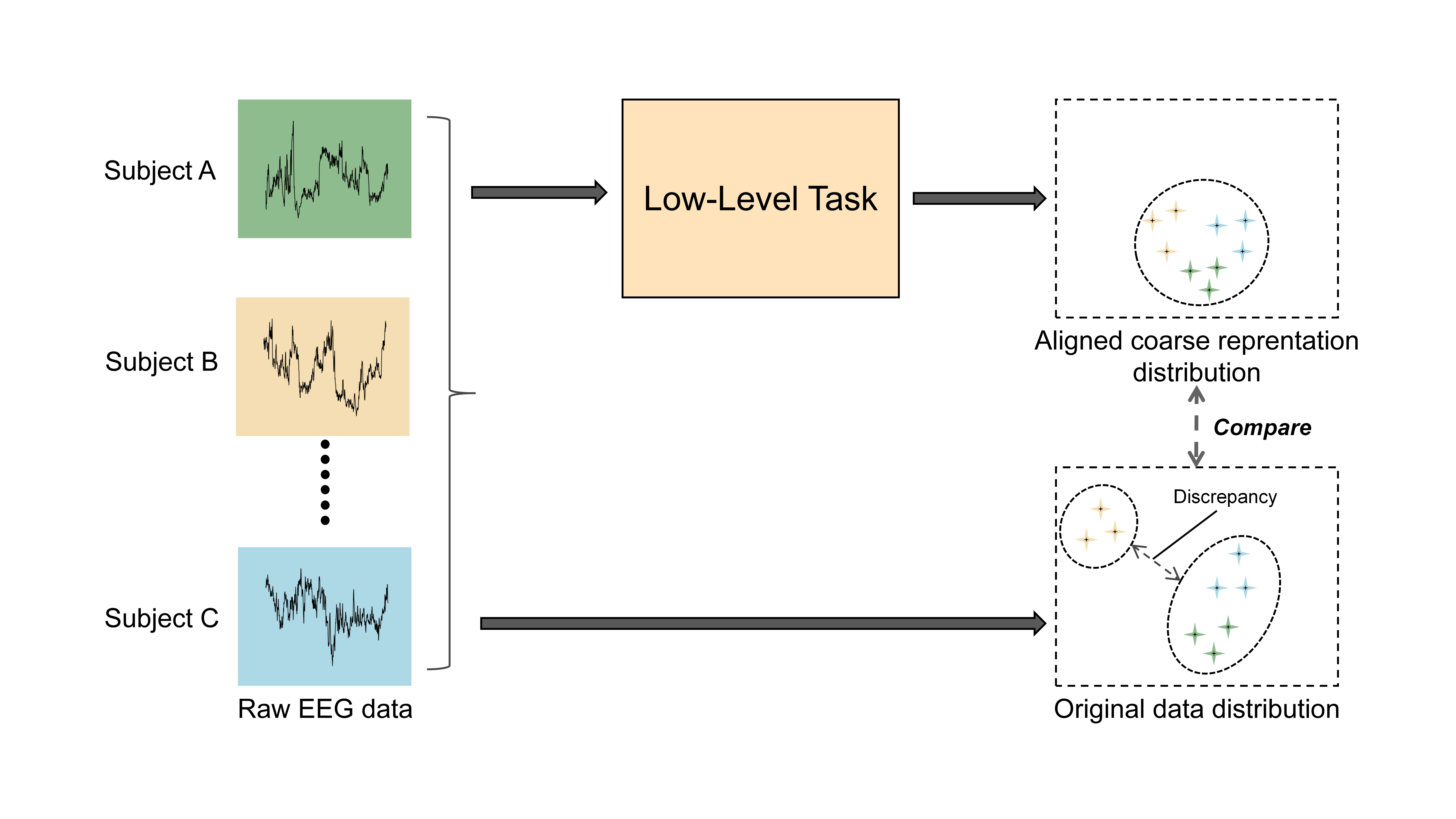}
  \caption{The demonstration of the effects of low-level tasks. The low-level task is expected to align the distribution and reduce the discrepancy between the data from various subjects.}
  \label{fig00}
\end{figure}

The time-to-frequency transform relationship provides a clue regarding how to define a low-level task for self-supervised EEG representation learning. A raw EEG signal in the time domain includes all the information to obtain the corresponding sample in the frequency domain. Thus, it is feasible to reconstruct the sample in the frequency domain using the representation learned from the time domain. To perform this reconstruction task, the model is actually forced to learn a Fourier-based transformation. Compared with the other self-supervised tasks, this task aims to learn a relatively coarse pattern involving less semantic information, i.e., a fixed transform relationship. In addition, this relationship holds invariant regardless of the subject from whom the EEG signal was collected. This fact provides a subject-invariant property of emotional EEG responses. By leveraging this property, we can align data from different subjects and map them into a subject-invariant distribution, thereby preserving more subject-invariant information for subject-independent emotion recognition. This inspiration motivates us to formalize such a relationship into a low-level task suitable for integration within the cascaded architecture.

To address these issues, this work proposes a cascaded self-supervised architecture to learn representation from emotional EEG signals. The purpose of our work is to explore the potential of the low-level task to facilitate the performance of the high-level self-supervised task. To this end, we formulate the relationship between time and frequency domain into a low-level reconstruction task, named time-to-frequency reconstruction (TFR). The representation learned in the time domain is used to reconstruct the input sample in the frequency domain. In addition, contrastive learning modules are adopted in our work as a high-level self-supervised learning task. Specifically, our proposal is based on a two-stream architecture. The proposed TFR module is connected to a contrastive learning module in the time-domain stream. Then, our work performs the second contrastive learning task in the frequency-domain stream to enhance the representation learning ability further. All three modules are combined in a joint loss to optimize this architecture. In the experiment section, we perform our method on two public benchmark datasets, DEAP and DREAMER. The comparison with existing methods demonstrates that our method outperforms similar works and shows a competitive performance when compared with the others. Besides that, our experiment also illustrates the indispensability of such a low-level task for performance by canceling or replacing the TFR module, further validating the effectiveness of our proposal. Moreover, the capacity of our method with limited labeled data is evaluated to assess the resistance of our method to label scarcity. 

Our primary contribution can be summarized as follows:

$\bullet$ We propose to cascade the low-level and high-level self-supervised task in a pipeline. Our proposed method follows a low-to-high representation learning procedure, aiming to preserve more generalizable representations. In this manner, the proposed method is expected to exhibit enhanced inter-subject generalizability on previously unseen data.

$\bullet$ To define an appropriate low-level task, we propose a time-to-frequency reconstruction task for the proposed low-to-high pipeline. The representation learned in time domain is used to reconstruct its frequency-domain features, learning the low-level representation.

$\bullet$ We implemented extensive experiments on two public benchmark datasets. Our results show that our method outperforms the existing similar works and show a competitive performance over the others. Besides that, our experiment further demonstrates the effectiveness of the proposed modules.

\section{Related works}
\subsection{Subject-independent EEG emotion recognition}
Considering the practicability, many researchers are interested in subject-independent approaches capable of recognizing emotions well across different subjects. However, EEG signal presents a high inter-subject variability. This characteristic of EEG signals makes it difficult for models to perform well when shifting from training data to testing data. Currently, there are two approaches to cope with this issue. One way is through domain adaptation (DA), which aims to reduce the differences between the training and test data. For example, Zheng et al.~\cite{r12} applied classical DA methods to the SEED dataset. In addition, domain-adversarial neural network~\cite{r13} (DANNs) is also introduced to this area, using a domain classifier to learn domain-indiscriminate representation. This method further led to ideas like bi-hemispheres domain-adversarial neural network~\cite{r10} (Bi-DANN) and regularized graph neural network~\cite{r11} (RGNN). Bi-DANN uses two hemisphere domain classifiers and a global domain classifier to get subject-independent emotion representations. RGNN changes the usual way of training to be more effective. These methods make the inter-subject model work better, but they rely on access to the test data. However, the test data is usually inaccessible in practice, limiting the application of EEG-based emotion recognition algorithms. Another way is domain generalization (DG), a promising method for subject-independent EEG emotion recognition. This method allows models to perform well across subjects without access to test data. DG aims to extract subject-invariant representation applicable to various subjects. Ma et al.~\cite{r14} extended the domain-adversarial neural networks (DANNs) into a (DG) method. They introduced a domain residual network (DResNet) that learns domain-shared and domain-specific weights. However, most of the existing subject-independent works still follow the supervised training strategy, limiting the practicability in real-world applications. Recently, a novel contrastive learning method, CLISA, as presented in~\cite{r7}, has been introduced for subject-independent EEG-based emotion recognition. CLISA couples two samples from distinct subjects, each corresponding to the same emotional stimuli, as anchor and positive samples. This method introduces a robust contrastive learning framework tailored for EEG-based emotion recognition, reaching a better performance than the work in~\cite{r14}. More importantly, this work draws further attention to subject-independent approaches in a self-supervised manner. 
\subsection{Self-supervised representation learning for EEG emotion recognition}


Self-supervised representation learning has exhibited notable achievements in many research fields, such as natural language processing, computer vision, etc. Recently, although the application of self-supervised learning in EEG emotion recognition still needs to be explored~\cite{r0}, we still have seen a few efforts adopting this training strategy in some works. For example, in~\cite{r15}, authors apply multiple signal transformations to the original signals and use the signal transformation recognition as the pretext task. And, the work presented in \cite{r2} proposes a self-supervised contrastive learning framework, using a genetics-inspired data augmentation method named meiosis. Moreover, the self-supervised reconstruction task is also adopted to learn the representation through a masked autoencoder architecture in~\cite{r16}. However, these works are subject-dependent algorithms. Considering the practicability of subject-independent models, some works aim to explore the subject-independent self-supervised learning in this area. The authors in~\cite{r7} propose a contrastive learning method named CLISA, which maximize the similarity of inter-subject EEG responses to the same emotional stimuli in the representation space. While their inspiring work demonstrates effectiveness in capturing inter-subject correlations, there remains room for further improvement in performance. In~\cite{r17}, a novel LSTM with attention mechanism is proposed to extract subject-invariant features of EEG data, based on self-supervised reconstruction pretext task. Although this work demonstrates impressive performance on public datasets, it also shows reliance on the well-design and complex network architecture. In the broader context of time-series research, TF-C, as introduced in~\cite{r21}, also recognizes the potential of time-frequency properties in facilitating self-supervised learning. However, our approach diverges from TF-C. TF-C utilizes the time-frequency property to make a novel definition for the different views of the data in contrastive learning. It assumes that the time-based and the frequency-based representations should show consistency in the latent space under the guidance of the proposed novel contrastive loss. However, we treat the time-frequency property as a clue to define the low-level self-supervised task. The newly formulated low-level task is expected to facilitate the finding of subject-invariant features by evolving the learning procedure from low-level to high-level. Consequently, our work should not be perceived as redundant or overlapping with TF-C. We compare the performance with TF-C in the experiment section to underscore this distinction.

\section{Method}
\begin{figure*}[ht]
  \includegraphics[width=\textwidth]{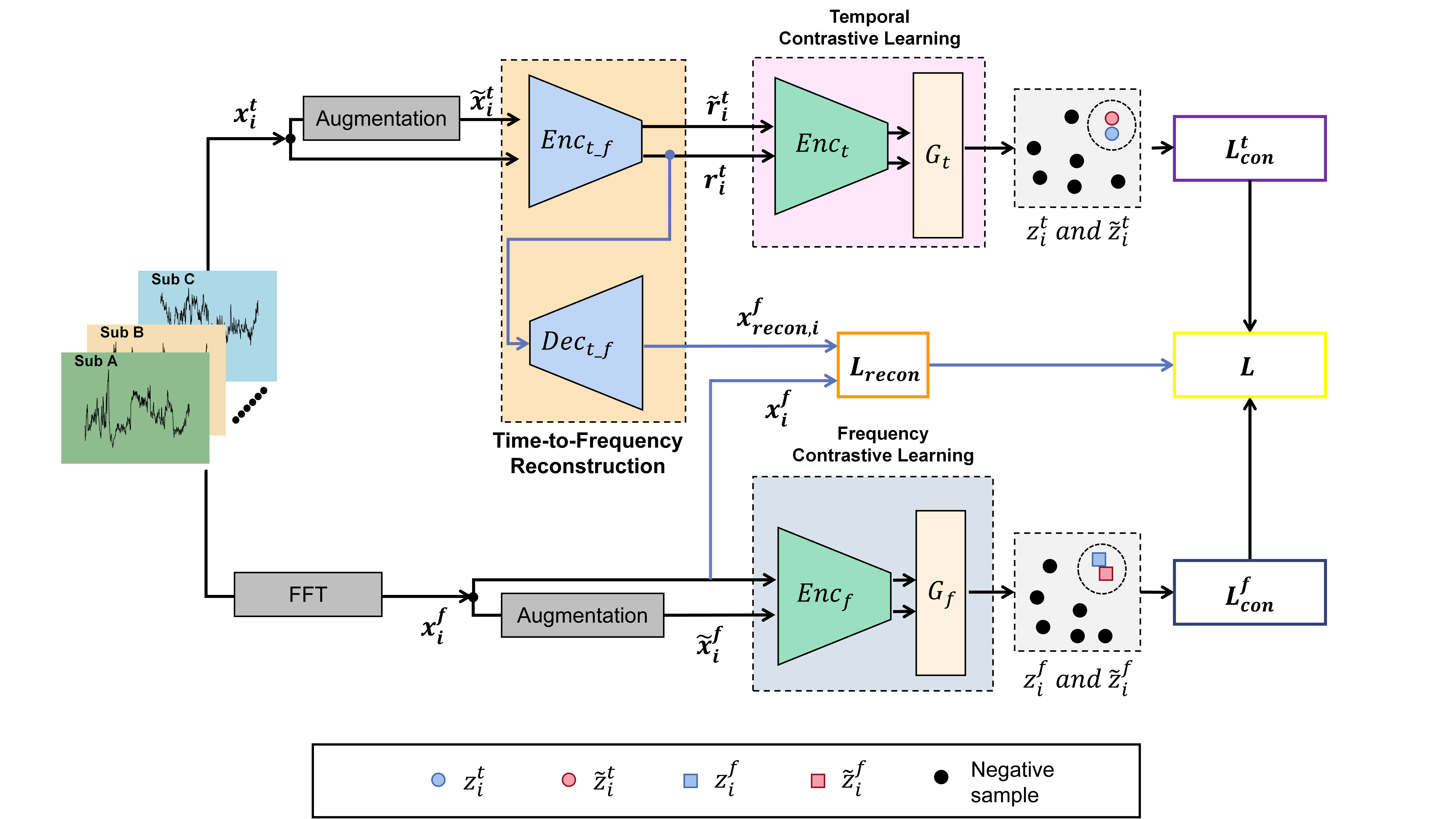}
  \caption{The overview of the representation learning framework. The samples that the should be put together are circled with dashed line. The input raw EEG sample $x_i^t$ is first transformed into the frequency domain by Fast Fourier Transform (FFT). Then, we generate augmentations $\tilde{x}^{t}_i$ and $\tilde{x}^{f}_i$ for $x^{t}_i$ and $x^{f}_i$. Subsequently, time-to-frequency reconstruction and temporal contrastive learning are performed in a cascaded pipeline. Moreover, the frequency contrastive learning is implemented for a more comprehensive representation learning. Finally, we sum the reconstruction loss $L_{recon}$ and two contrastive learning losses $L_{con}^t$ and $L_{con}^f$ as the final loss $L$.}
  \label{fig1}
\end{figure*}

\subsection{Overview}
In this section, the cascaded self-supervised learning architecture will be described in detail as shown in Fig~\ref{fig1}. First, a low-level self-supervised task is performed at the beginning. The input of EEG data is projected into the frequency domain via the fast Fourier Transform. Next, the EEG data is sent to a depth-wise convolution layer to learn EEG representation in the time domain, and a linear layer is added to reconstruct the EEG data in the frequency domain from the learned representation. Then, two contrastive learning modules, i.e., temporal and frequency contrastive learning, are embedded to implement the high-level contrastive learning task. In particular, temporal contrastive learning adopts the learned representation in the TFR task as input. Instead, the frequency contrastive learning adopts the raw EEG frequency spectrum as input. Finally, all three tasks are combined to optimize the model in a joint training process. 

\subsection{Time-to-frequency reconstruction}

As we mentioned above, a time-to-frequency reconstruction task is formulated as a low-level self-supervised task. To this end, we propose an encoder-decoder structure to reconstruct the input sample in the frequency domain, using the representation extracted from the input sample in the time domain. This objective is supposed to force the encoder to learn the time-to-frequency transform relationship, preserving the subject-invariant information. Let $\{{x}_i^t|{x}_i^t\in R^{C\times T}\}^N_1, $ denote a batch of raw EEG data with size $N$ in the time domain. Moreover, we use $\{{x}^f|{x}_i^f\in R^{C\times T}\}^N_1$ to represent the generated frequency spectrum. Our model learn the representation $r_i^t$ in time domain can be seen in the Eq~\ref{eq1}

\begin{equation}
    \label{eq1}
    \begin{aligned}
    &r_i^t=Enc_{t\_f}(x_i^t), r_i^t\in R^{C\times T}, \\
    &x^f_{recon,i}=Dec_{t\_f}(r_i^{t}), 
    \end{aligned}
\end{equation}
where the $x^f_{recon, i}$ denotes the reconstructed input in the frequency domain, $Enc_{t\_f}$ denotes the encoder, and the $Enc_{t\_f}$ denotes the decoder for TFR.

In order to encourage the consistency between the reconstruction $x^f_{recon}$ and the original frequency-domain input ${x
}^f$, we adopt the mean square error (MSE) to measure the similarity between $x^f_{recon}$ and ${x}^f$ as shown in Eq~\ref{eq2}.
\begin{equation}
    \label{eq2}
    L_{recon}=\frac{1}{N}\sum^N_i||x^f_{recon,i}-{x}_i^f||_2^2,
\end{equation}
where $L_{recon}$ denotes the reconstruction loss.

\subsection{Temporal and frequency contrastive learning}
To further enhance the effectiveness of extracted representation, we include contrastive learning tasks following the TFR. The strategy of contrastive learning is to align the representations between the different views of data. Thus, the first step is to create the augmentation for the original EEG data. Here, we adopt the augmentation methods in~\cite{r21} for both time and frequency domains, generating the augmentation sets $\{\tilde{x}^t_i\}_1^N$ and $\{\tilde{x}^f_i\}_1^N$. In both domains, we randomly select an augmenting method for each sample from the bank. The time-domain augmentation bank includes jittering, scaling, time shift, and neighborhood segmentation. And the frequency-domain augmentation bank includes removing and adding frequency components. Then, two encoders are set to learn the representation in time and frequency domains, respectively.

\begin{equation}
    \label{eq3}
    \begin{aligned}
    &h_i^t=Enc_t(r_i^t),  \tilde{h}_i^t=Enc_t(\tilde{r}_i^t),\\
    &h_i^f=Enc_f(x_i^f),  \tilde{h}_i^f=Enc_f(\tilde{x}_i^f),
    \end{aligned}
\end{equation}
where $\tilde{r}_i^t=Enc_{t\_f}(\tilde{x}_i^t)$, $Enc_t$ denotes the time-domain encoder, and $Enc_f$ denotes the frequency-domain encoder. 

Subsequently, the learned representations are projected into a latent space for alignment. For this reason, two projectors, $G_t$ and $G_f$, are added into the pipeline following the $Enc_t$ and $Enc_f$. 
\begin{equation}
    \label{eq5}
    \begin{aligned}
    &z_i^t=G_t(h_i^t),  \tilde{z}_i^t=G_t(\tilde{h}_i^t),\\
    &z_i^f=G_f(h_i^f),  \tilde{z}_i^f=G_f(\tilde{h}_i^f),
    \end{aligned}
\end{equation}  
where $z_i^t$ represents the projection of $h_i^t$ in the latent space, with analogous definitions for $\tilde{z}_i^t$, $z_i^f$, and $\tilde{z}_i^f$. 

Then, we construct the contrastive losses to guide the optimization. Initially, we merge the two sets $\{z_i^t\}^N_1$ and $\{\tilde{z}_i^t\}^N_1$ into a new sets $\{\overline{z}_i^t\}^{2N}_1$ with size of $2N$. Similarly, we also merge the $\{z_i^f\}^N_1$ and $\{\tilde{z}_i^f\}^N_1$ into $\{\overline{z}_i^f\}^{2N}_1$. Notably, $\overline{z}_{2k-1}^t$ and $\overline{z}_{2k}^t$ in the new set, $k\in \{1,\dots,N\}$, are regarded as positive for each other because they are different views of the $z^t_k$, and so are the cases of $\{\overline{z}_i^f\}^{2N}_1$. Let $I=\{1,\dots,2N\}$
 denotes the indexes set. The $p^t(i)$ denotes the index of the positive sample of $\overline{z}^t_i$, and $p^f(i)$ denotes the index of the positive sample of $\overline{z}^f_i$. Subsequently, the contrastive learning losses can be calculated as described in Eq~\ref{eq6}.
\begin{equation}
    \label{eq6}
    \begin{aligned}
    &L_{con}^t=-\sum_{i\in I}log\frac{exp(z_i^t\cdot z_{p^t}^t(i)/\tau)}{\sum_{a\in A(i)} exp(z_i^t\cdot z_a^t/\tau)},\\
    &L_{con}^f=-\sum_{i\in I}log\frac{exp(z_i^f\cdot z_{p^f}^f(i)/\tau)}{\sum_{a\in A(i)} exp(z_i^t\cdot z_a^t/\tau)},
    \end{aligned}
\end{equation}
where $A(i)=\{a|a\in I, a\neq i\}$, and $\tau$ is  a scalar temperature parameter. We use $L_{con}^t$ to denote the loss in time domain and $L_{con}^f$ to denote the loss in frequency domain.
\subsection{Optimization and prediction}

\begin{figure}[ht]
  \centering
  \includegraphics[width=\linewidth]{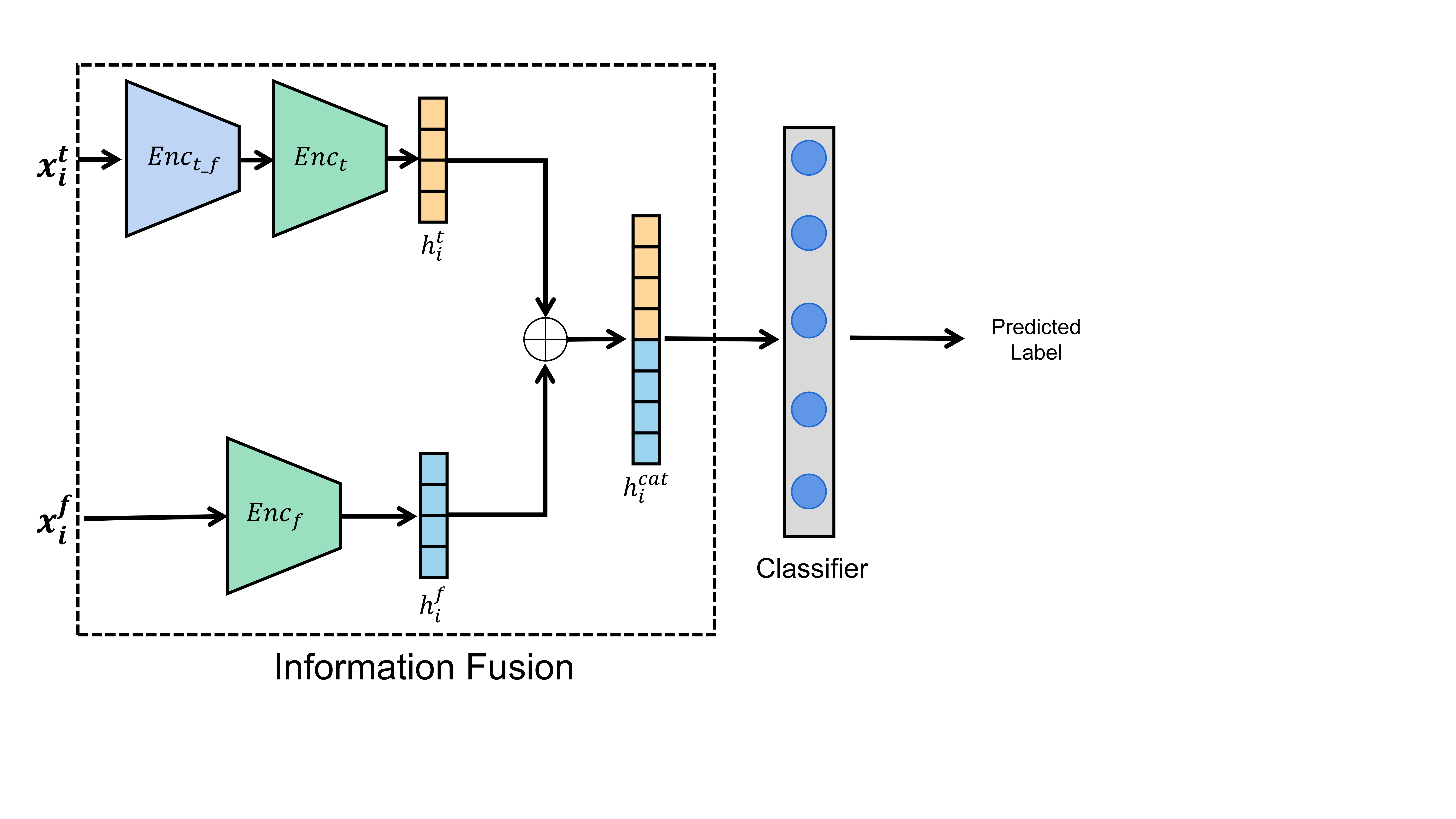}
  \caption{The overview of prediction procedure. The $\oplus$ denotes the concatenating operation. We concatenate the representations from various encoders together to fuse the information in the both streams. Finally, the classifier makes decision based on the concatenated representation $h^{cat}_i$.}
  \label{fig2}
\end{figure}

The final representation learning loss comprises three components: temporal contrastive loss, frequency contrastive loss, and time-to-frequency reconstruction loss as described in~\ref{eq7}.
\begin{equation}
    \label{eq7}
    L=\lambda(L^t_{con}+L^f_{con})+(1-\lambda)L_{recon},
\end{equation}
where $\lambda$ is a hyper-parameter that controls the balance of contrastive and reconstruction losses. 

In the prediction phase, the trained model is refined through a fine-tuning process on unseen test data. Augmentations, decoders, and projectors are excluded, retaining only the three trained encoders to extract representations $h^t_i$ and $h_i^f$, followed by the flattening operation. These representations are then concatenated to achieve information fusion, resulting in the final representation $h^{cat}_i$. Subsequently, a classifier is trained to classify the learned final representations into various emotion status. The whole process is visualized in Fig~\ref{fig2}. 

\subsection{Detailed structure}
The structure of the encoders, decoder, non-linear projectors, and the classifier are presented in details in this subsection. Considering the property of Fourier-based transform, we adopted a depth-wise 1-D convolution layer as the encoder structure to learn representation from ${x}_i^t$ for TFR. And then, a linear layer is adopted as a decoder to reconstruct ${x}_i^f$.

\begin{equation}
    \label{eq8}
    \begin{aligned}
    &Enc_{t\_f}(x_i^t)=Conv1D_{dep}(x_i^t;K_1,S_1,P_1), \\
    &Dec_{t\_f}(r_i^{t})=r_i^{t}W, W\in R^{T\times T},
    \end{aligned}
\end{equation}
where $K_1$ denotes the kernel, $S_1$ denotes the stride, and $P_1$ denotes the paddding.

In this work, we aim to present a compact encoder structure suited to our task for contrastive learning. For simplicity, the two encoders in both domains, $Enc_t$ and $Enc_f$, adopt the identical structure. For fewer trainable parameters, our work adopts separable one-dimensional 2-D convolutions to learn the pattern on different dimensions separately~\cite{r7,r22}. First, we extend the input size into $1\times C\times T$. Next, we adopt a convolution layer with kernel size $(1, K_2)$ to learn the temporal or frequency pattern. Moreover, Existing research demonstrates that the responses to emotion between the right and left hemispheres of the brain show an asymmetric pattern~\cite{r24}. Considering that, we follow the~\cite{r23} to adopt global and hemisphere kernels, learning the spatial pattern. In detail, we adopt two convolution layers with kernel size $(C/2,1)$ and $(C,1)$. Finally, the outputs of the two spatial convolution layers are concatenated along the spatial dimension and fused by a one-dimensional 2-D convolution layer with kernel size $(K_3\times 1)$. The learning process of our encoder structure can be described in the Eq~\ref{eq4}.
\begin{equation}
    \label{eq4}
    \begin{aligned}
    &h=Enc(x); \\
    &h_1=Conv2D^{tf}(x;S_1,P_2,F_1),\\
    &h_{21}=Conv2D^{spa}_{glb}(h_1;S_1,F_2),\\
    &h_{22}=Conv2D^{spa}_{hem}(h_1;S_2,F_2),\\
    &h_2=Concat(h_{21},h_{22}), h_2\in R^{C\times T},\\
    &h_{fu}=Conv2D^{fuse}(h_2),\\
    &h=AvgPool(LReLU(h_{fu})).
    \end{aligned}
\end{equation}
The $Conv2D^{tf}$ denotes the convolution layer that learns temporal or frequency pattern, the $Conv2D^{spa}_{glb}$ denotes the convolution layer that learns the global spatial pattern, the $Conv2D^{spa}_{hem}$ denotes the convolution layer that learns the hemisphere spatial pattern, and $Conv2D^{fuse}$ denotes the convolution layer that fuse the outputs of $Conv2D^{spa}_{glb}$ and $Conv2D^{spa}_{hem}$. Besides, $S_1$ and $S_2$ represent the strides, $P_3$ represents the padding, and $F_2$ represents the filter size. Finally, $AvgPool$ represents an average pooling operation with kernel $(1,K_4)$ and stride $S_3$, and $LReLU$ represents the LeakyReLU activation layer.

As for the projector, we adopt a non-linear projector structure for both contrastive learning modules. It comprise of two fully-connected layers with a batch normalization layer and a ReLUs layer inserted in the middle. The hidden dimension of the two fully-connected layers are 256 and 128. And we build a three-layer multilayer perceptron (MLP) and used it as a classifier.The MLP consists of two hidden layers, each with 30 units. Rectified linear units (ReLUs) are inserted between every two layers.

\section{Experiment}
\subsection{Dataset and Preprocessing}
Our experiments are conducted on two widely used benchmark datasets: DEAP and DREAMER. DEAP~\cite{r25} is a multi-modal dataset focused on human affective computing, consisting of 32 subjects, each participating in 40 trials. Subjects watch one-minute music videos, self-reporting their emotional states in arousal and valence dimensions. For our experiment, we adopt the 32-channel EEG signals included in it that are recorded at 512 Hz during the trial. Then, the EEG signals is down-sampled to 128 Hz. This experiment approach this as a binary classification task, transforming the 9-level labels into low and high classes. Additionally, following~\cite{r23}, we segment the trial into 4-second segments.

DREAMER~\cite{r26} features recordings of 14-channel EEG signals captured during affect-inducing audio-visual stimuli. With 23 subjects watching 18 movie clips, the duration ranging from 63 to 393 seconds. Participants are asked to rate arousal and valence using self-assessment manikins (SAM) scores from 1 to 5. Similar to DEAP, we categorize the 5-level labels into low and high classes. We employ sliding windows to segment EEG recordings, breaking each trial into 9-second segments with a 1-second slide.

\subsection{Implementation details}
We empirically select hyperparameters for the implementation, as detailed in Table~\ref{tab1}. In the representation learning procedure, we set $\tau$ to 0.07. All subsequent experiments are conducted using an NVIDIA RTX 2080Ti GPU. The representation learning model employs Adam as the optimizer, with a learning rate of $0.0001$ for the DEAP dataset and $0.00008$ for the DREAMER dataset. The batch size is configured as 128. For the classifier, we also use the Adam optimizer to optimize its cross-entropy loss with a learning rate of $0.00001$.

\begin{table}[th]
  \centering
  \caption{Hyperparameter Settings}
  \label{tab1}
  \begin{tabular}{|c|c|c|c|}
    \toprule
    Hyper-Parameter&Value&Hyper-Parameter&Value\\
    \midrule
    $K_1$& 25&$S_3$& (1,4)\\
    $K_2$ & 49&$P_1$ & 12\\
    $K_3$ & 3&$P_2$ & (0,24)\\
    $K_4$ & 4&$F_1$ & C*C\\
    $S_1$ & 1&$F_2$ & 16*16\\
    $S_2$ & (C/2,1)&$\lambda$ &0.1\\

  \bottomrule
  \end{tabular}
\end{table}

\subsection{Performance Evaluation Protocol}

We implement the proposed method on the DEAP and DREAMER datasets. To assess the inter-subject generalization of our approach, we employ the leave-one-subject-out protocol for evaluation. This protocol reserve one subject for evaluation, utilizing the remainder of subjects for training. This process is repeated across all subjects in the dataset. As the result, the mean accuracy and standard deviation is calculated to measure the subject-independent performance.

Significantly, for enhanced practicality, we aspire for our model to exhibit commendable performance even in the absence of access to test data. For this reason, both the representation learning modules and the classifier are exclusively trained on the training data. It is noteworthy that our proposed model refrains from fine-tuning on the test data, positioning it as a subject-independent work comparable with domain generalization methods.

\subsection{Comparison with the Existing Methods}
\begin{table}[h]
    \centering
    \caption{The Mean Accuracy and Standard Deviation of Existing Emotion Recognition Models on DEAP}
    \label{tab2}
    \begin{threeparttable}
    \begin{tabular}{c|c|c|c}
        \hline
        &Method& Arousal  & Valence \\
        \hline
        \multirow{11}*{Supervised} & TSVM~\cite{r12}   & 56.59$\pm$11.98 & 61.77$\pm$8.93\\
        &TPT~\cite{r12} & 54.76 $\pm$12.48      & 57.43$\pm$14.54 \\
        &TCA~\cite{r12} & 51.81$\pm$15.03     & 56.23 $\pm$14.33\\
        &KPCA~\cite{r12} & 58.15 $\pm$14.96     & 54.35 $\pm$10.22\\
        &RODAN~\cite{r27}&56.60$\pm$3.48&56.78$\pm$3.3\\
        &AD-TCN~\cite{r28}&63.25$\pm$ 4.62 &64.33$\pm$7.06 \\
        &Wang et al.~\cite{r29}&69.79$\pm$11.93&66.47$\pm$8.75\\
        &BiSMSM*~\cite{r30}&61.87$\pm/$ &62.97$\pm/$ \\
        &VMD*~\cite{r31}&61.25$\pm/$&62.50$\pm/$\\
        &TSception*~\cite{r23} & 63.67$\pm$10.30    & 60.26$\pm$6.51 \\
        &DeepConvNet*~\cite{r32} & 63.39$\pm$9.74    & 60.22$\pm$6.13 \\
        &ShallowConvNet*~\cite{r32}  & 61.37$\pm$10.93    & 59.85$\pm$6.07 \\
        \hline
        \multirow{3}*{Self-supervised} &CLISA*~\cite{r7}&64.50$\pm$10.1&61.46$\pm$6.7\\
        &EEGFuseNet*~\cite{r33}&58.55$\pm/$&56.44$\pm/$\\
        &TF-C*~\cite{r21}&63.00$\pm$12.85&59.23$\pm$7.97\\
        &\textbf{Ours*}&\textbf{69.67$\pm$8.85}&\textbf{65.50$\pm$5.47}\\
        \hline
    \end{tabular}
    \begin{tablenotes}
        \footnotesize
        \item[*] The method without the need of access to target domain.
    \end{tablenotes}
    \end{threeparttable}
\end{table}

Here, our method is compared with three groups of methods: supervised subject-independent, self-supervised subject-independent, and intra-subject. To begin with, the supervised subject-independent methods comprise DA and DG methods, with which we will compare our method. In the first place, the classical DA baseline methods, i.e., TSVM, TPT, TCA, and KPCA, are compared to ours. Besides that, we also compare our work with some recent DA approaches in this area. The approaches outlined in~\cite{r27,r28} employ the widely recognized adversarial training strategy to minimize dissimilarities between diverse subjects. Additionally, in~\cite{r29}, the authors introduce a Symmetric and Positive Definite (SPD) matrix network (daSPDnet) aimed at capturing shared emotional representations with short calibration. These endeavors represent recent advancements in DA for our specific task. Notably, these methods necessitate access to test data for optimal performance. In contrast, our methodology aligns with DG principles, obviating the need for test data access. Simultaneously, we conduct comparative evaluations with alternative DG methods~\cite{r30,r31}. In~\cite{r31}, the approach focuses on extracting invariant features through Variational Mode Decomposition (VMD). Moreover,~\cite{r30} proposes capturing discriminative features from multiple perspectives, demonstrating commendable performance across various subjects.

However, these supervised works need the label information to guide the representation learning. In~\cite{r7,r33}, the authors explore self-supervised learning for subject-independent EEG emotion recognition. Besides,~\cite{r21} proposes a self-supervised contrastive learning using time-frequency consistency for time-series pre-training. Here, we transfer it to our task for comparison. Specifically, we re-implement the CLISA and TF-C using the reported setting in~\cite{r7,r21}. For a fair comparison, we performed a segment-level classification rather than a trial-level classification in~\cite{r7}. Besides that, we cancel the fine-tuning process, training the TF-C architecture on the train data only. 

Recently, there have been many effective methods designed for within-subject EEG emotion recognition. These approaches have shown their ability to capture discriminative information for identifying human emotions in EEG signals. Consequently, there is a theoretical potential for these methods to perform inter-subject task. To explore this, we compare our method with some representative methods, e.g., DeepConvNet~\cite{r32}, ShallowConvNet~\cite{r32}, and TSception~\cite{r23}. All the within-subject methods are evaluated following the leave-one-subject-out protocol. The comparison helps highlight how well our method generalizes to different individuals.

The experiment is conducted on two widely used benchmark datasets, namely DEAP and DREAMER. The outcomes for the DEAP dataset are presented in Table~\ref{tab2}. Our proposed method exhibits markedly better performance compared to existing self-supervised methods, highlighting its efficacy in the self-supervised learning domain. In terms of a comparison with supervised subject-independent models, our method outperforms all the listed methods without requiring access to test data. While not universally surpassing all the presented DA methods, our approach remains competitive and generally outperforms a significant portion of them. These experimental findings substantiate the robustness of our method in capturing subject-invariant representations as a self-supervised method.
\begin{table}[ht]
    \centering
    \caption{The Mean Accuracy and Standard Deviation of Existing Emotion Recognition Models on DREAMER}
    \label{tab3}
    \begin{threeparttable}
    \begin{tabular}{c|c|c|c}
        \hline
        &Method& Arousal  & Valence  \\
        \hline
        \multirow{8}*{Supervised}&TSVM~\cite{r12}& 55.67$\pm$12.07& 60.76$\pm$9.77 \\
        &TPT~\cite{r12} & 61.89$\pm$13.18    & 59.22$\pm$15.01 \\
        &TCA~\cite{r12}  & 54.37$\pm$8.56    & 55.85$\pm$6.45 \\
        &KPCA~\cite{r12}  & 60.03$\pm$11.24    & 53.74$\pm$8.47 \\
        &AD-TCN~\cite{r28}&63.69$\pm$6.57&66.56$\pm$10.04\\
        &Wang et al.~\cite{r29}&76.57$\pm$14.04&67.99$\pm$6.34\\
        &BiSMSM*~\cite{r30}&61.87$\pm/$ &62.97$\pm/$ \\
        &TSception*~\cite{r23} & 62.60$\pm$8.16    & 64.19$\pm$8.48 \\
        &DeepConvNet*~\cite{r32} & 65.84$\pm$7.35    & 65.88$\pm$6.81 \\
        &ShallowConvNet*~\cite{r32}  & 64.58$\pm$6.50    & 63.61$\pm$7.45 \\
        \hline
        \multirow{2}*{Self-supervised}&CLISA*~\cite{r7}&62.14$\pm$10.03 &63.04$\pm$8.83\\
        &TF-C*~\cite{r21}&60.95$\pm$12.99& 62.65$\pm$10.56\\
        &\textbf{Ours*}&\textbf{71.04$\pm$6.06}&\textbf{69.63$\pm$7.07}\\
        \hline
    \end{tabular}
    \begin{tablenotes}
        \footnotesize
        \item[*] The method without the need of access to target domain.
    \end{tablenotes}
    \end{threeparttable}
\end{table}

Furthermore, the experimental outcomes on the DREAMER dataset are detailed in Table~\ref{tab3}. Notably, our approach maintains its superior performance over all self-supervised methods, mirroring the observations on the DEAP dataset. Compared to supervised methods, our model remains competitive and outperforms most of them. While our model is slightly outperformed by the method in~\cite{r29} in the arousal dimension on mean accuracy, this discrepancy is justifiable due to the lack of access to label information and test data. This outcome underscores the efficacy of our approach in enhancing generalizability within a self-supervised framework. In summary, our method demonstrates superior performance compared to similar approaches and remains competitive against alternative methods.

\subsection{Effectiveness of the Proposed Time-to-frequency Reconstruction}

To further investigate the effectiveness of the proposed TFR, we perform experiments on DREAMER using two variants of the proposed method. In the first variant, we replace the TFR with a conventional time-to-time reconstruction task. Specifically, the decoder is modified to reconstruct the original features in the time domain. The $L_{recon}$ in Eq~\ref{eq2} is changed to $L_{recon}=\frac{1}{N}\sum^N_i||x^t_{recon,i}-{x}_i^t||2^2$, where $x^t_{recon,i}$ denotes the output of the decoder. With this variant, we aim to verify the effectiveness of the proposed low-level reconstruction task compared to the other reconstruction task. In the second variant, we aim to testify the capacity of proposed model without any reconstruction task. For this reason, we remove the decoder while retaining the encoder in the TFR module. In this way, the variant cancels the low-to-high representation learning procedure but maintains a similar learning capacity to perform the subsequent contrastive learning module. The results can be seen in Table~\ref{tab4}.
\begin{table}[ht]
  \centering
  \caption{The Performance on Different Training Data Scales}
  \label{tab4}
  \begin{tabular}{|c|c|c|} \hline  
    
    Variant&Arousal&Valence \\ \hline
    Proposed method&71.04$\pm$6.06&69.63$\pm$7.07\\ \hline
    Proposed method w/ t-t recon&70.75$\pm$6.53&67.89$\pm$7.86\\ \hline
    Proposed method w/o recon&70.34$\pm$6.48&67.61$\pm$8.07\\ \hline
  
  \end{tabular}
  \begin{tablenotes}
        \footnotesize
        \item w/o recon: without any reconstruction
        \item w/ t-t recon: with a conventional time-to-time reconstruction task
    \end{tablenotes}
\end{table}

As we can see, the TFR module plays an indispensable role in enhancing the performance of the proposed model, outperforming all modified variants in both dimensions. Furthermore, when considering the second variant as a baseline, it is obvious that the efficacy of the conventional time-to-time reconstruction, a high-level self-supervised task, is limited in comparison to the proposed TFR. This strengthen the indispensability of a low-level task to boost performance. In conclusion, the experimental results support the efficacy of the proposed TFR.

\subsection{Performance with Limited Labeled Data}
\begin{figure}[ht]
  \centering
  \includegraphics[width=\linewidth]{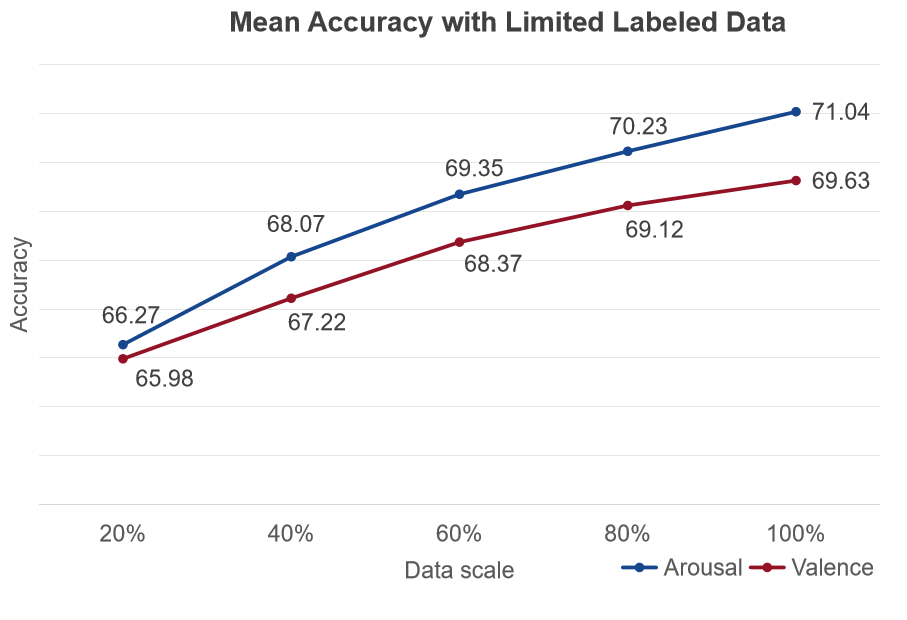}
  \caption{The mean accuracy of the proposed method with limited labeled data on DREAMER benchmark dataset.}
  \label{fig3}
\end{figure}

Given the scarcity of labeled data, it is imperative to assess the performance of our method with limited training data. Consequently, we systematically reduce the labeled data to percentages of $20\%$, $40\%$, $60\%$, and $80\%$. In order to maintain data balance, we assign the same pre-defined proportions of reserved labeled data for each subject in each trial. Subsequently, the classifier is trained using the reserved labeled data, while the representation learning architecture is trained using all the data. And we still adopt leave-one-subject out protocol to evaluate the performance. The experiment is conducted on the DREAMER benchmark dataset, and the results are presented in Fig~\ref{fig3}.

\subsection{Visualization}

\begin{figure}[ht]
  \centering
  \includegraphics[width=\linewidth]{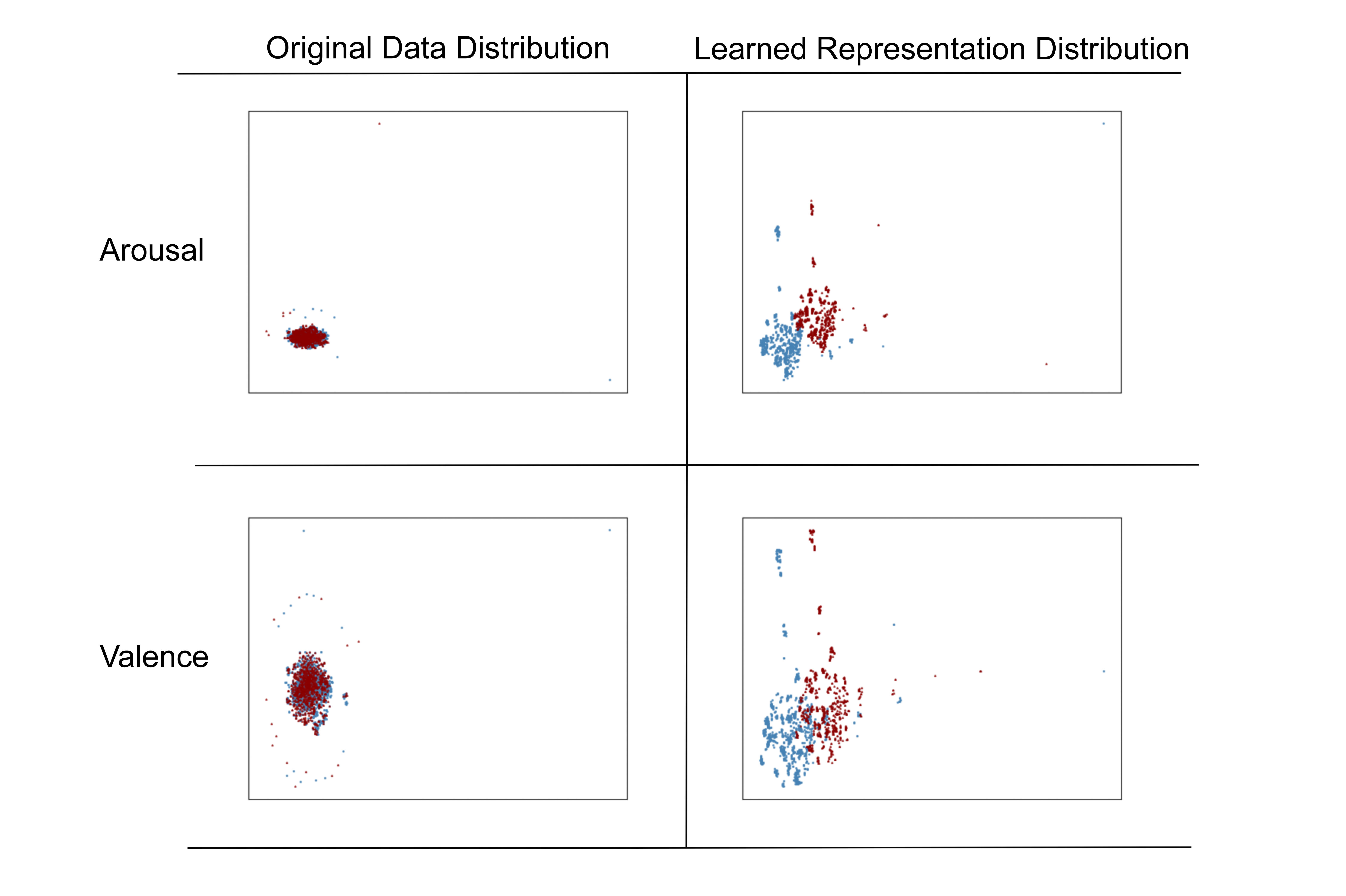}
  \caption{The visualization of the learned representation and the original EEG data for all subjects. The red points correspond to samples labeled as 1, and the blue points correspond to samples labeled as 0.}
  \label{fig5}
\end{figure}

To provide a more intuitive demonstration of the effect of our model, we employ the T-SNE (t-distributed stochastic neighbor embedding) technique to visualize the data in the DREAMER dataset in a two-dimensional space. Initially, to show the capacity of our model to learn subject-independent discriminative emotional information, we randomly sampled one-tenth of the raw data from each subject for analysis. As shown in Fig~\ref{fig5}, both the learned representations and the original EEG data are visualized for comparison. In the distribution of the original data, different emotional states are almost inseparable. In contrast, in the distribution of learned representations, different emotional states could be separated more effectively. This observation supports the effect of our model on learning subject-independent emotional information.

\begin{figure}[ht]
  \centering
  \includegraphics[width=\linewidth]{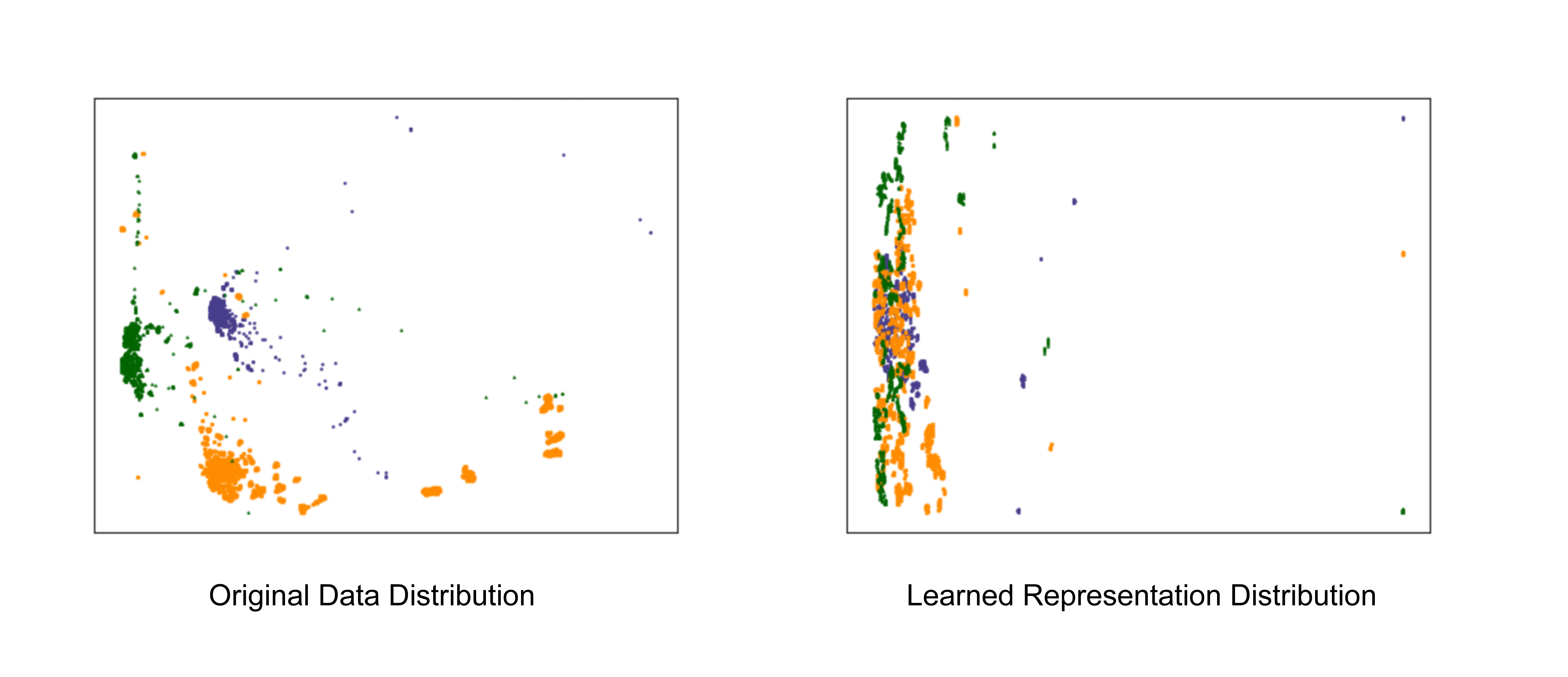}
  \caption{The visualization of the learned representation and the original EEG data for three randomly selected subjects. The data points with various colors corresponds to the data collected from various subjects.}
  \label{fig6}
\end{figure}

Besides that, we further demonstrate the capacity of our model to reduce the inter-subject discrepancy. In Fig~\ref{fig6}, we randomly select three subjects and visualize their data. The points of original data collected from the same subject tend to cluster together, and the data clusters of different subjects tend to show an obvious discrepancy. In contrast, in the right part of Fig~\ref{fig6}, the data points of learned representation from various subjects are merged. It indicates that the inter-subject discrepancy is reduced through the proposed learning process.

\begin{figure}[ht]
  \centering
  \includegraphics[width=\linewidth]{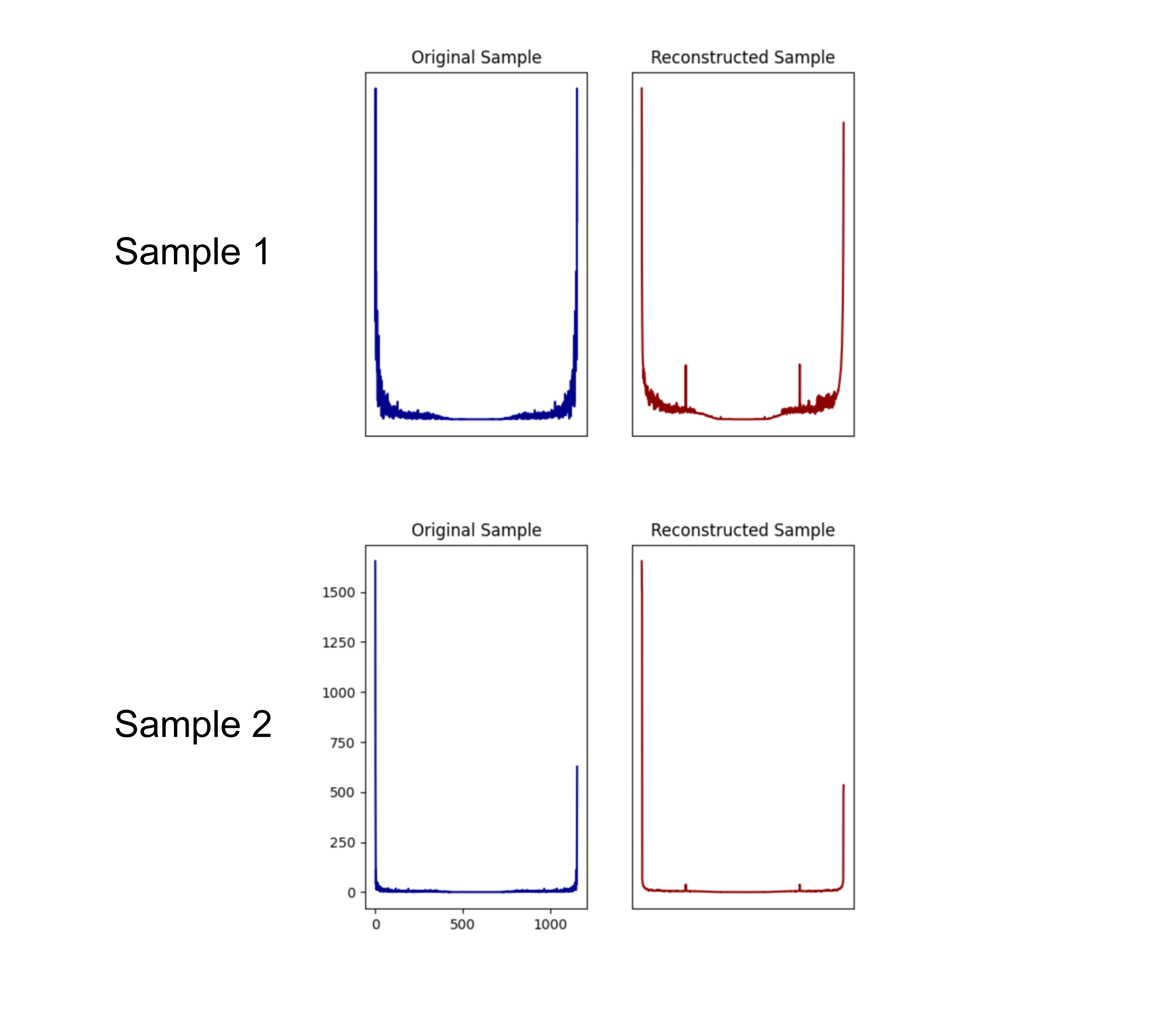}
  \caption{The visualization of the original EEG data in the frequency domain and the reconstructions.}
  \label{fig7}
\end{figure}

To enhance the understanding of the proposed TFR task, we employ the T-SNE technique to illustrate its functionality. Initially, we randomly select multiple EEG signals from various subjects and channels as an example, ensuring the findings are representative. And then, the selected frequency-domain original samples alongside their corresponding reconstructions are visualized as shown in Fig~\ref{fig7}. Although the generated reconstruction results have some errors compared to the original data, the reconstructions still preserve the basic characteristics of the original signals. Therefore, we can assume that the TFR task has succeeded in learning a mapping relation that approximates the Fourier-based transform, preserving enough subject-invariant information and aligning the data distribution.

\subsection{Ablation Study}
To further assess the efficacy of the components within our proposed method, we conduct an ablation study on the arousal dimension of the DREAMER dataset. This subsection outlines three variants of our method, aiming to discern the impact of each component on the overall performance of the model. The experiments adhere to the leave-one-subject-out protocol. The three variants are delineated as follows.

$\bullet$\textbf{Single time-domain stream}. In this variant, we remove the encoder in the frequency-domain stream and the time-to-frequency reconstruction. And, the input of classifier $h_i^{cat}$ is replaced with $h_i^{t}$. Our goal is to examine the necessity of the complementary information in the frequency-domain stream and the proposed TFR module.

$\bullet$\textbf{Base model}. To further prove the effectiveness of our method, we are interested in the learning capacity of the encoder and classifier without the low- and high-level tasks. In this variant, we combine the encoder and the classifier into an end-to-end deep learning model. Without the two-stream architecture, we only use the raw EEG sample as the input of the new base model. Because the projector is designed to learn the mapping to a suitable latent space, it is not necessary anymore when we remove the contrastive learning modules. Hence, we also discard the projector in this variant. The representation learned from the raw EEG sample is directly flattened and input into the classifier. The combination of the encoder and the classifier is regarded as a whole structure optimized by the cross-entropy loss.

$\bullet$\textbf{Time-to-frequency reconstruction}. In earlier subsection, we discuss the efficacy of the TFR module. However, the performance of the model only consists solely of a TFR module has not yet been evaluated. In order to obtain a more reliable demonstration, we are also interested in exploring the representation learning capacity of the low-level TFR module without the help of high-level task. For this reason, we cancel all the components in the learning framework, only reserving the TFR module to extract the representation.  

\begin{figure}[ht]
  \centering
  \includegraphics[width=\linewidth]{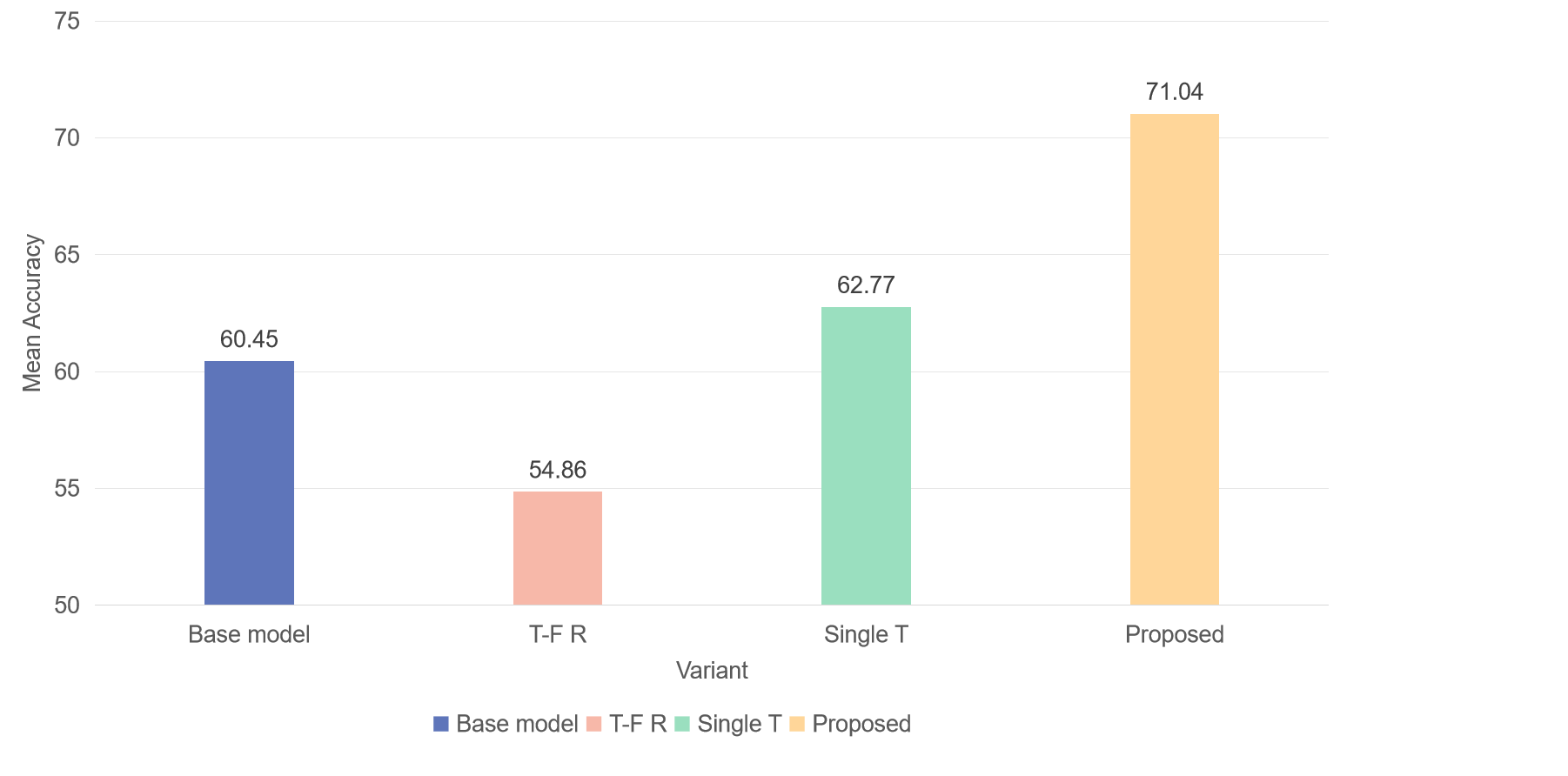}
  \caption{The mean accuracy results of the variants in the ablation study.}
  \label{fig4}
\end{figure}

The results can be seen in the Fig~\ref{fig4}. All the variants suffer a decline in performance in different degrees. It proves that the proposed modules have the effect of improving subject-independent performance. Notably, the Time-to-frequency reconstruction variant reaches the worst mean accuracy, 54.86\%. This observation suggests that a low-level task is insufficient to solely capture the key information for emotion recognition, needing the refinement of the following high-level tasks.

\section{Conclusion}
In this work, we investigate the effectiveness of cascaded low-to-high architecture in enhancing the generalizability of self-supervised learning for emotion EEG recognition. To perform such an architecture, we define a novel time-to-frequency reconstruction task as a low-level self-supervised task. Moreover, we incorporate contrastive learning into the proposed architecture as a high-level task. Our extensive experiments demonstrate that the proposed self-supervised learning method can reach an advanced performance compared with the existing methods. Besides that, our further experiments highlight the indispensability of such a low-level task of our model by evaluating the model's performance when the TFR module is replaced or canceled. In summary, our work substantiates its proposals in enhancing the generalizability of the self-supervised EEG-based emotion recognition model.

In addition, we also proposes to suggest avenues for future exploration. There remains room to advance the low-to-high paradigm in this area. Future research can delve into the alternative types of low-level self-supervised tasks. Moreover, the integration of low-level and high-level tasks is a promising avenue for future advance. The relationship between the two tasks in depth is still under-explored, which might expose the reasons underlying their collective impact on the final learned representation. This motivation can be further extended to investigate the design of interaction mechanisms between different tasks.

\section*{References}

\bibliographystyle{IEEEtran}
\bibliography{main}
\end{document}